# Structural study of a rare earth-rich aluminoborosilicate glass containing various alkali and alkaline-earth modifier cations

A. Quintas, D. Caurant, O. Majérus, M. Lenoir
*Laboratoire de Chimie de la Matière Condensée de Paris, UMR 7574, ENSCP, Paris, France*

A. Quintas, J-L. Dussossoy
*Commissariat à l'Energie Atomique, Centre d'étude de la Vallée du Rhône,
DEN/DTCD/SCDV/LEBV, Bagnols-sur-Cèze, France*

T. Charpentier
*CEA Saclay, Laboratoire de Structure et Dynamique par Résonance Magnétique,
DSM/DRECAM/SCM – CEA CNRS URA 331, Gif-sur-Yvette, France*

D. Neuville
*Laboratoire de Physique des minéraux et des Magmas, UMR 7047-CNRS-IPGP, Université Pierre et Marie Curie, Paris, France,*

A rare-earth rich aluminoborosilicate glass of composition (given in wt.%): 50.68 $SiO_2$ – 4.25 $Al_2O_3$ - 8.50 $B_2O_3$ – 12.19 $M_2O$ – 4.84 M'O – 3.19 $ZrO_2$ – 16.35 $Nd_2O_3$ (where M and M' are respectively an alkali and alkaline earth cation) is currently under study as potential nuclear waste form. In this work, we were interested in the structure of this glass in relation with the modifier cation type. Two different glass series were elaborated by changing separately the nature of the alkaline (M=Li, Na, K, Rb, Cs) and the alkaline-earth (M'=Mg, Ca, Sr, Ba) ions and different structural studies were intended to elucidate the local environment of the rare-earth and the network arrangement. Only slight effect was put in evidence on the covalency degree and the length of Nd-O linkage with a change of M or M', by optical spectroscopy and EXAFS measurements. Raman and MAS NMR ($^{29}$Si, $^{27}$Al, $^{11}$B) spectroscopies showed a variation of the polymerization degree of the network with the size of the modifier cation. Finally, the most important feature of this glass composition is related to the $AlO_4^-$ charge compensation which was proved to be uniquely assured by alkali cations.

## Introduction

A new confinement glass, aimed at immobilizing more concentrated nuclear waste solutions, is currently under study. A rare-earth (RE) rich aluminoborosilicate glass [1,2] has already proved, from a technological point of view (chemical durability, thermal stability and waste capacity), to be good candidate for the immobilization of these high level radioactive wastes. Its simplified seven-oxides glass composition (referred to as glass A) is 50.68 $SiO_2$ – 4.25 $Al_2O_3$ - 8.50 $B_2O_3$ – 12.19 $Na_2O$ – 4.84 CaO – 3.19 $ZrO_2$ – 16.35 $Nd_2O_3$ (wt.%). In this system, sodium and calcium are supposed to simulate respectively all the other alkali and alkaline-earth cations present in the complete glass composition and originating from both waste solutions and the glass frit. In former studies, special attention was laid on the role of both modifier cations sodium and calcium [3-5] by studying glasses with different Ca/(Na+Ca) ratios. However, the effect of other modifier cation type on the glass structure is still a matter of discussion (especially in such particular complex system) and requires further investigations. In this work, we are interested in the structure of this glass in relation with the modifier cation type. For this purpose, two different glass series were elaborated: an alkali glass series in which the nature of the alkali is successively changed (Li, Na, K, Rb and Cs) and an alkaline-earth glass series in which the alkaline earth cation is also changed (Mg, Ca, Sr and Ba). Different structural studies were intended to elucidate the local environment of the neodymium ions and the network arrangement in order to reach global comprehension of the glass structure.

## Results and discussion

Optical spectroscopy at 10K and EXAFS measurements (at the Nd $L_{III}$ edge) revealed only slight effect on the covalency degree and the length of the Nd-O linkage even though significant change in the modifier cationic field strength is produced by the substitution. This indicates that the neodymium sites are quite well-defined and suggests that the $Nd^{3+}$ ions are able to maintain their appropriate environment in the glass composition studied in this work.

$^{11}$B MAS NMR spectroscopy revealed significant modification of the B(III)/B(IV) ratio with changing the modifier cation type. In the alkali glass series, the relative B(IV) content is strongly enhanced when the alkaline cationic field strength increases (except for the lithium bearing glass which gives way to a lower B(III)/B(IV) ratio than the sodium bearing glass). This suggests a better affinity of $BO_4$ species for smaller alkali cations. On the contrary, in the alkaline-earth glass series, the impact of the nature of the alkaline-earth cation is very weak (except for the magnesium bearing glass which exhibits lower B(IV) concentration). $^{27}$Al MAS NMR spectroscopy allowed to elucidate the nature of aluminium (IV) charge compensation in our glasses. According to the spectra presented in figure 1, changing the nature of the alkali cation generates a strong increase in the quadrupolar coupling constant whereas strictly no modification is observed when changing the nature of the alkaline-earth cation. This demonstrates that $AlO_4^-$ entities are almost exclusively balanced by alkali cations.

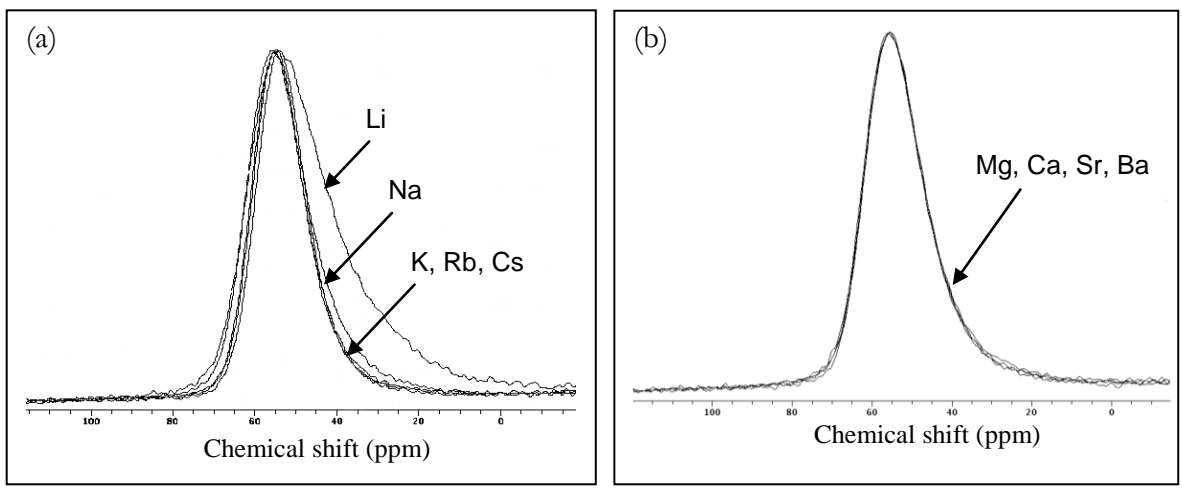

Figure 1 : $^{27}$Al MAS NMR spectra recorded for the alkaline series (a) and alkaline-earth series (b), normalized to the same intensity ($B_0$=11.75T, $\nu_{rot}$=12.5kHz).

Raman spectra of the high frequency region displayed in figure 2 exhibit significant variation in the glass polymerization which may be due to a change in the NBOs distribution in the glass network. This evolution is consistent with the $Q^3 \leftrightarrow Q^4+Q^2$ disproportionation reaction displaced towards the right with increasing cationic field strength. The particularly strong variation observed for the lithium bearing glass is expected to be due to the occurrence of glass in glass phase separation as already observed in other glasses of similar composition [5].

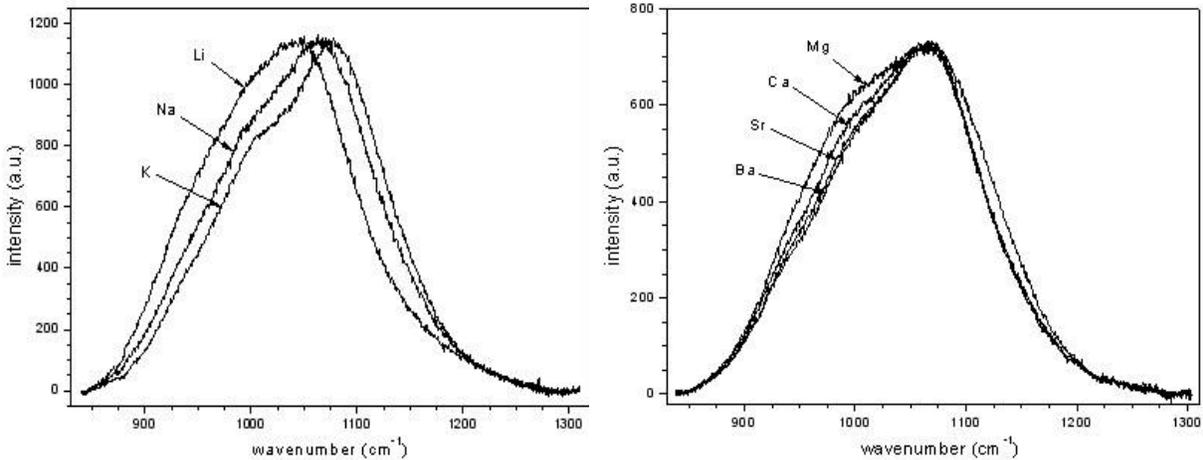

Figure 2 : Normalized Raman spectra of the high frequency region for the alkaline series (a) and alkaline-earth series (b), corrected with the Long formula (excitation source at λ=488nm).

**Conclusion**

In this aluminoborosilicate glass it was shown that the neodymium environment is only slightly affected by a change in the modifier cation type which means that the high field strength rare earth, located in highly depolymerized region of the glass network, is able to maintain its appropriate environment. It was also put in evidence a better affinity of $BO_4^-$ species for smaller alkali cations. Finally, it was established that the negative charge of the $AlO_4^-$ species, in this glass composition, is almost uniquely balanced by alkali cations.

1 I. Bardez, D. Caurant, P. Loiseau, J-L. Dussossoy, C. Gervais, F. Ribot, D. R. Neuville, N. Baffier, *Phys. Chem. Glasses* **46**, 3120 (2006).

2 I. Bardez, D. Caurant, J-L. Dussossoy, P. Loiseau, C. Gervais, F. Ribot, D.R. Neuville, N. Baffier, C. Fillet, *Nucl. Sci. Eng.* **153**, 272 (2006).

3 A. Quintas, O. Majérus, D. Caurant, J-L. Dussossoy, P. Vermaut, *J. Am. Ceram. Soc.* **90**, 712 (2007).

4 A. Quintas, O. Majérus, M. Lenoir, D. Caurant, K. Klementiev, A. Webb, J-L. Dussossoy, *J. Non-Cryst. Solids* **354**, 98 (2008).